\documentclass[fleqn,usenatbib]{mnras}

\usepackage{newtxtext,newtxmath}

\usepackage[T1]{fontenc}

\DeclareRobustCommand{\VAN}[3]{#2}
\let\VANthebibliography\thebibliography
\def\thebibliography{\DeclareRobustCommand{\VAN}[3]{##3}\VANthebibliography}

\usepackage{graphicx}	
\usepackage{amsmath}	

\title[GRB 221009A: Secondary emission from UHECRs]{Secondary GeV-TeV emission from ultra-high-energy cosmic rays accelerated by GRB 221009A}

\author[N. Mirabal]{
N. Mirabal,$^{1}$\thanks{E-mail: nestor.r.mirabalbarrios@nasa.gov}
\\
$^{1}$Mail Code 661, Astroparticle Physics Laboratory, NASA Goddard Space Flight Center,
Greenbelt, MD 20771, USA\\
$^{2}$University of Maryland, Baltimore County, MD 21250, USA\\
$^{3}$Center for Research and Exploration in Space Science and Technology, NASA Goddard
Space Flight Center, Greenbelt, MD 20771\\
}

\date{Accepted XXX. Received YYY; in original form ZZZ}

\pubyear{2015}

\begin{document}
\label{firstpage}
\pagerange{\pageref{firstpage}--\pageref{lastpage}}
\maketitle

\begin{abstract}
The origin of ultra-high-energy cosmic rays (UHECRs) remains elusive. Gamma-ray bursts (GRBs) are among the best candidates able to meet the stringent energy requirements needed for particle acceleration to such high energies. If UHECRs were accelerated by the central engine of GRB 221009A, it might be possible to detect secondary photons and neutrinos  as the UHECRs travel from the source to the Earth. Here we attempt to interpret some of the early publicly available data connected to this burst. If the reported early GeV-TeV detection was produced by secondary emission from UHECRs it probably indicates that UHECRs reached energies  $> 10^{21}$ eV and that GRB 221009A exploded inside a magnetic void with intergalactic magnetic field (IGMF) strength  $B \leq 3 \times 10^{-16}$ G.  In order to understand the entire energy deposition mechanism, we propose to search existing and future Fermi-LAT data  for secondary emission arriving over larger spatial scales and longer time-scales. This strategy might help clarify the origin of UHECRs, constrain the intergalactic magnetic field (IGMF) strength along this line of sight and start to quantify the fraction of magnetic voids around GRBs.
\end{abstract}




\section{Introduction}
GRBs are thought to be plausible sites for UHECR acceleration \cite{waxman2,vietri}. The recent detection of GRB 221009A \citep{swift,gbm,lat,pillera} is a local reminder of the tremendous energy release that is associated with GRBs.  Located at  
a distance of 721 Mpc with a redshift z = 0.1505 \citep{2022GCN.32648....1D,gcn32686}, it is most likely the most energetic GRB ever detected \citep{2022GCN.31365....1A}. Unfortunately, GRB 221009A lies beyond the GZK distance horizon and UHECRs from the explosion are unlikely to reach us directly. 
However, if UHECRs were accelerated as part of the GRB outflow, secondary emission above 100 GeV is expected either from UHECRs cascading into secondary photons  \cite{1996ApJ...464L..75W,mirabal2}, or from ultra-high-energy photons (produced by UHECR cascades near the GRB site) spilling out of the host galaxy and later cascading in magnetic void regions  \citep{2012ApJ...745L..16M}. Potential evidence for such secondary emission has been presented around the nearby GRB 980425/SN 1998bw and discussed for GRB 221009A \citep{mirabal1,mirabal2}.
Here we expand on this work and also propose to continue searching for additional secondary emission  from GRB 221009A with the Fermi Large Area Telescope (LAT) on larger spatial scales and longer time-scales.

\section{Relevant Equations and Scales}\label{relevant}

Once UHECRs initiate pair cascades,  the relevant time-scale for secondary emission delay $\tau$ is dictated by the delay of electron-positron pairs  in a magnetic field of strength $B$ \citep{1996ApJ...464L..75W,mirabal2}. 

\begin{equation}
 \tau \sim (10^{3} - 10^{5}) B^2_{-11}~{\rm yr},
\end{equation}

\noindent
corresponding to the high-energy end $E_{e} = 10^{18}$ eV and  low energy end $E_{e} = 10^{15}$ eV of the cascade. In this equation $B$~might correspond to either the magnetic field within the host galaxy or the IGMF depending on the location of the UHECRs along its trajectory to Earth.

The LHAASO detector array has reported the detection of $>$ 500 GeV emission within 2000~s after the GRB 221009A trigger \citep{gcn32677}. Just as impressive, the Carpet-2 air shower array has also reported an air-shower consistent with a 251-TeV photon energy 4536~s after the GBM trigger \citep{2022ATel15669....1D}. If  GeV-TeV photons were created of part of UHECR cascades with a time delay $\tau \leq 2000$ s, an IGMF strength $B \leq 3 \times 10^{-16}$ G along the line of sight would be required. Typically, the time it takes UHECRs to cross $\sim$ kpc scales in  GRB host galaxy with $\mu$G  magnetic fields is of order $\sim $ days \citep{1996ApJ...464L..75W}. This can be significantly shortened to scale of $\sim$ hours if UHECRs have energies $> 10^{21}$ eV. The complete delay equations can be found in \citet{mirabal2}.

It is conceivable that
the early TeV emission could have originated in UHECR cascades but there are three conditions for that to work: 1) UHECR energies  $> 10^{21}$ eV in order to be able to cross the host galaxy in less than one hour and 2) the GRB explosion must have taken place inside a magnetic void with $B \leq 2 \times 10^{-16}$ G in order to avoid additional time delays in the IGMF, and 3) TeV photons at this distance must be able to survive absorption in the EBL. Although meeting these conditions appear to be rather stringent, it seems plausible. 

Remarkably, this scenario would be able to explain ultra-high energy cosmic rays to astounding energies reaching $E > 10^{21}$ eV neatly. We must mention that there is an alternative  scenario that would not require UHECRs to cross the host galaxy in very short time-scales. In the latter, ultra-high-energy photons are produced by UHECR cascades near the GRB site, these ultra-high energy photons subsequently spill out of the host galaxy and later cascade in the surrounding magnetic void  \citep{2012ApJ...745L..16M}. A GRB explosion taking place inside a magnetic void appears unavoidable in both scenarios. 

The IGMF value implied from our analysis is rather low and as already mentioned points to the presence of a magnectic void.
Generally, a detection of secondary emission from nearby GRBs in human timescales is thought to require  IGMF strengths of order $B \leq 10^{-13}$ G \citep{mirabal1,mirabal2}. 
However, no two GRBs will happen under the same magnetic field conditions. By continuing monitoring additional secondary emission with the Fermi LAT, we should be able to gauge the IGMF strength along different lines of sight.

\section{Optimized LAT Analysis Strategy}\label{strategy}
Depending on what type of nuclei were formed in the GRB 221009A outflow, the angular size of the secondary emission could be quite extended \citep{2011MNRAS.415.2495M}. Fortunately, 
the exceptional all-sky coverage and nearly uninterrupted observations afforded by the Fermi LAT makes a detection of even more extended secondary emission feasible \citep{mirabal1,mirabal2}. Extended secondary emission from a much wider region should be searched for in existing and future LAT data for GRB 221009A. 
Since the secondary emission does not have a well-formed spatial model, a standard point-source likelihood fit with the Fermi LAT would not work straight away. Initially, we suggest using aperture photometry centered at the GRB 221009A position  \footnote{https://fermi.gsfc.nasa.gov/ssc/data/analysis/scitools/aperture\_photometry.html}. One possibility is to start with a 1-degree radius as the initial aperture and search for additional secondary emission incrementally outwards {\it e.g.} $2^{n}$-degree radii for $n \leq 3$. Since GRB 221009A is located near the Galactic plane, the other obvious issue is contamination by unrelated foreground/background flares. To deal with this, we propose to use two distinct energy ranges (signal and background). For instance, the
100 GeV - 4 TeV energy range could be taken as the signal band since most of the secondary emission is expected at $> 0.1$ TeV energies. The 1 GeV-100 GeV energy range  would then set the background in order to rule out unrelated fluctuations including Galactic transients (novae) and extragalactic transients (AGN).

\section{Conclusions}\label{discussion}
The origin of UHECRs is a hotly contested subject. The recent detection of early TeV emission in GRB 221009A might be potentially explained if UHECR energies reached $> 10^{21}$ eV and the GRB took place within a magnetic void with   $B \leq 3 \times 10^{-16}$ G.  Searches for secondary emission on larger spatial scales and longer time-scales are encouraged. As stated earlier, a major source of uncertainty throughout is the EBL attenuation. First considerations  point out that certain EBL models could be transparent enough to allow a TeV detection at a $z=0.1505$ \citep{2022arXiv221010778Z}, but more detailed EBL studies are needed. Another possibility is that UHECRs escaped the host galaxy and initiated the cascades at a much lower redshift to ease the EBL constraints. A LHAASO spectrum above 500 GeV might help determine directly the EBL attenuation associated with GRB 221009A.  High-energy neutrinos could also have been produced in such a powerful UHECR accelerator \citep{1998PhRvD..59b3002W}. Unfortunately, an early non-detection has been reported by the IceCube collaboration \citep{icecube22}. It appears that the next generation of UHE neutrino detectors such as IceCube2, Trinity, and GRAND will be needed for a significant detection in real time  \citep{2022arXiv221015625M,2022arXiv221014116A}. 
As already shown with GRB 980425/SN 1998bw \citep{mirabal1,mirabal2}, progress in our understanding of UHECRs might have finally arrived by leveraging observations of secondary emission along the line of sight to nearby GRBs. 

\section*{Acknowledgements}
The material is based upon work supported by NASA under award number 80GSFC21M0002. We acknowledge useful correspondence with Kohta Murase.  
We thank the referee for useful comments that helped improve the paper.

\section*{Data Availability}

All data used in this paper are public.



\bibliographystyle{mnras}
\bibliography{References} 

\begin{thebibliography}{}
\makeatletter
\relax
\def\mn@urlcharsother{\let\do\@makeother \do\$\do\&\do\#\do\^\do\_\do\%\do\~}
\def\mn@doi{\begingroup\mn@urlcharsother \@ifnextchar [ {\mn@doi@}
  {\mn@doi@[]}}
\def\mn@doi@[#1]#2{\def\@tempa{#1}\ifx\@tempa\@empty \href
  {http://dx.doi.org/#2} {doi:#2}\else \href {http://dx.doi.org/#2} {#1}\fi
  \endgroup}
\def\mn@eprint#1#2{\mn@eprint@#1:#2::\@nil}
\def\mn@eprint@arXiv#1{\href {http://arxiv.org/abs/#1} {{\tt arXiv:#1}}}
\def\mn@eprint@dblp#1{\href {http://dblp.uni-trier.de/rec/bibtex/#1.xml}
  {dblp:#1}}
\def\mn@eprint@#1:#2:#3:#4\@nil{\def\@tempa {#1}\def\@tempb {#2}\def\@tempc
  {#3}\ifx \@tempc \@empty \let \@tempc \@tempb \let \@tempb \@tempa \fi \ifx
  \@tempb \@empty \def\@tempb {arXiv}\fi \@ifundefined
  {mn@eprint@\@tempb}{\@tempb:\@tempc}{\expandafter \expandafter \csname
  mn@eprint@\@tempb\endcsname \expandafter{\@tempc}}}

\bibitem[\protect\citeauthoryear{{Ai} \& {Gao}}{{Ai} \&
  {Gao}}{2022}]{2022arXiv221014116A}
{Ai} S.,  {Gao} H.,  2022, arXiv e-prints, \href
  {https://ui.adsabs.harvard.edu/abs/2022arXiv221014116A} {p. arXiv:2210.14116}

\bibitem[\protect\citeauthoryear{{Atteia}}{{Atteia}}{2022}]{2022GCN.31365....1A}
{Atteia} J.~L.,  2022, GRB Coordinates Network, \href
  {https://ui.adsabs.harvard.edu/abs/2022GCN.31365....1A} {31365, 1}

\bibitem[\protect\citeauthoryear{{Bissaldi}, {Omodei}  \& M.}{{Bissaldi}
  et~al.}{2022}]{lat}
{Bissaldi} E.,  {Omodei} N.,   M. K.,  2022, GRB Coordinates Network, 32637, 1

\bibitem[\protect\citeauthoryear{{Castro-Tirado} et~al.,}{{Castro-Tirado}
  et~al.}{2022}]{gcn32686}
{Castro-Tirado} A.~J.,  et~al., 2022, GRB Coordinates Network, 32686, 1

\bibitem[\protect\citeauthoryear{{Dzhappuev} et~al.,}{{Dzhappuev}
  et~al.}{2022}]{2022ATel15669....1D}
{Dzhappuev} D.~D.,  et~al., 2022, The Astronomer's Telegram, \href
  {https://ui.adsabs.harvard.edu/abs/2022ATel15669....1D} {15669, 1}

\bibitem[\protect\citeauthoryear{{Huang}, {Hu}, {Chen}, {Zha}, {Liu}  \&
  {Yao}}{{Huang} et~al.}{2022}]{gcn32677}
{Huang} Y.,  {Hu} S.,  {Chen} S.,  {Zha} M.,  {Liu} C.,   {Yao} Z.,  2022, GRB
  Coordinates Network, 32677, 1

\bibitem[\protect\citeauthoryear{{Icecube Collaboration}}{{Icecube
  Collaboration}}{2022}]{icecube22}
{Icecube Collaboration} 2022, GRB Coordinates Network, 32665, 1

\bibitem[\protect\citeauthoryear{{Kennea} \& {Williams}}{{Kennea} \&
  {Williams}}{2022}]{swift}
{Kennea} J.~A.,  {Williams} M.,  2022, GRB Coordinates Network, 32635, 1

\bibitem[\protect\citeauthoryear{{Metzger}, {Giannios}  \&
  {Horiuchi}}{{Metzger} et~al.}{2011}]{2011MNRAS.415.2495M}
{Metzger} B.~D.,  {Giannios} D.,   {Horiuchi} S.,  2011, \mn@doi [\mnras]
  {10.1111/j.1365-2966.2011.18873.x}, \href
  {https://ui.adsabs.harvard.edu/abs/2011MNRAS.415.2495M} {415, 2495}

\bibitem[\protect\citeauthoryear{{Mirabal}}{{Mirabal}}{2022a}]{mirabal1}
{Mirabal} N.,  2022a, arXiv e-prints, \href
  {https://ui.adsabs.harvard.edu/abs/2022arXiv221010822M} {p. arXiv:2210.10822}

\bibitem[\protect\citeauthoryear{{Mirabal}}{{Mirabal}}{2022b}]{mirabal2}
{Mirabal} N.,  2022b, arXiv e-prints, \href
  {https://ui.adsabs.harvard.edu/abs/2022arXiv221011430M} {p. arXiv:2210.11430}

\bibitem[\protect\citeauthoryear{{Murase}}{{Murase}}{2012}]{2012ApJ...745L..16M}
{Murase} K.,  2012, \mn@doi [\apjl] {10.1088/2041-8205/745/2/L16}, \href
  {https://ui.adsabs.harvard.edu/abs/2012ApJ...745L..16M} {745, L16}

\bibitem[\protect\citeauthoryear{{Murase}, {Mukhopadhyay}, {Kheirandish},
  {Kimura}  \& {Fang}}{{Murase} et~al.}{2022}]{2022arXiv221015625M}
{Murase} K.,  {Mukhopadhyay} M.,  {Kheirandish} A.,  {Kimura} S.~S.,   {Fang}
  K.,  2022, arXiv e-prints, \href
  {https://ui.adsabs.harvard.edu/abs/2022arXiv221015625M} {p. arXiv:2210.15625}

\bibitem[\protect\citeauthoryear{{Pillera}, {Bissaldi}, {Omodei}, {La Mura}  \&
  {Longo}}{{Pillera} et~al.}{2022}]{pillera}
{Pillera} R.,  {Bissaldi} E.,  {Omodei} N.,  {La Mura} G.,   {Longo} F.,  2022,
  The Astronomer's Telegram, \href
  {https://ui.adsabs.harvard.edu/abs/2022ATel15656....1P} {15656, 1}

\bibitem[\protect\citeauthoryear{{Veres}, {Burns}, {Bissaldi}, {Lesage}  \&
  {Roberts}}{{Veres} et~al.}{2022}]{gbm}
{Veres} P.,  {Burns} E.,  {Bissaldi} E.,  {Lesage} S.,   {Roberts} O.,  2022,
  GRB Coordinates Network, 32636, 1

\bibitem[\protect\citeauthoryear{{Vietri}}{{Vietri}}{1995}]{vietri}
{Vietri} M.,  1995, \mn@doi [\apj] {10.1086/176448}, \href
  {https://ui.adsabs.harvard.edu/abs/1995ApJ...453..883V} {453, 883}

\bibitem[\protect\citeauthoryear{{Waxman}}{{Waxman}}{1995}]{waxman2}
{Waxman} E.,  1995, \mn@doi [\apjl] {10.1086/309715}, \href
  {https://ui.adsabs.harvard.edu/abs/1995ApJ...452L...1W} {452, L1}

\bibitem[\protect\citeauthoryear{{Waxman} \& {Bahcall}}{{Waxman} \&
  {Bahcall}}{1998}]{1998PhRvD..59b3002W}
{Waxman} E.,  {Bahcall} J.,  1998, \mn@doi [\prd] {10.1103/PhysRevD.59.023002},
  \href {https://ui.adsabs.harvard.edu/abs/1998PhRvD..59b3002W} {59, 023002}

\bibitem[\protect\citeauthoryear{{Waxman} \& {Coppi}}{{Waxman} \&
  {Coppi}}{1996}]{1996ApJ...464L..75W}
{Waxman} E.,  {Coppi} P.,  1996, \mn@doi [\apjl] {10.1086/310090}, \href
  {https://ui.adsabs.harvard.edu/abs/1996ApJ...464L..75W} {464, L75}

\bibitem[\protect\citeauthoryear{{Zhao}, {Zhou}  \& {Wang}}{{Zhao}
  et~al.}{2022}]{2022arXiv221010778Z}
{Zhao} Z.-C.,  {Zhou} Y.,   {Wang} S.,  2022, arXiv e-prints, \href
  {https://ui.adsabs.harvard.edu/abs/2022arXiv221010778Z} {p. arXiv:2210.10778}

\bibitem[\protect\citeauthoryear{{de Ugarte Postigo} et~al.,}{{de Ugarte
  Postigo} et~al.}{2022}]{2022GCN.32648....1D}
{de Ugarte Postigo} A.,  et~al., 2022, GRB Coordinates Network, \href
  {https://ui.adsabs.harvard.edu/abs/2022GCN.32648....1D} {32648, 1}

\makeatother
\end{thebibliography}





\bsp	
\label{lastpage}
\end{document}